\pdfoutput=1

\documentclass[11pt,twoside,a4paper,cmspaper,final,collab]{cms-tdr}
\def\svnVersion{462362P}\def\cmsCernNoTag{CERN-EP-2018-028}\def\cmsCernDate{\today}\def\cmsMessage{Published in the Journal of High Energy Physics as \href{http://dx.doi.org/10.1007/JHEP05(2018)148}{\doi{10.1007/JHEP05(2018)148.}}}

\begin{document}\cmsNoteHeader{EXO-17-011}

\newcommand{\HNl}{\ensuremath{N_l}\xspace}
\newcommand{\HNe}{\ensuremath{N_e}\xspace}
\newcommand{\HNnu}{\ensuremath{N_\nu}\xspace}
\newcommand{\WR}{\PWR\xspace}

\hyphenation{had-ron-i-za-tion}
\hyphenation{cal-or-i-me-ter}
\hyphenation{de-vices}
\RCS$Revision: 462236 $
\RCS$HeadURL: svn+ssh://svn.cern.ch/reps/tdr2/papers/EXO-17-011/trunk/EXO-17-011.tex $
\RCS$Id: EXO-17-011.tex 462236 2018-05-29 18:22:22Z gnegro $

\newlength\cmsTabSkip\setlength{\cmsTabSkip}{1ex}

\newcommand{\gR}{\ensuremath{g_\mathrm{R}}}
\newcommand{\gL}{\ensuremath{g_\mathrm{L}}}
\newcommand{\WRv}{\ensuremath{\PW_{\mathrm{R}}^*}}
\newcommand{\NR}{\ensuremath{\mathrm {N_R}}\xspace}
\newcommand{\mNR}{\ensuremath{m_{\NR}}}
\newcommand{\mWR}{\ensuremath{m_{\WR}}}
\newcommand{\mll}{\ensuremath{m_{\ell\ell}}}
\newcommand{\mlljj}{\ensuremath{m_{\ell\ell \text{jj}}}}
\newcommand{\meejj}{\ensuremath{m_{\Pe\Pe \text{jj}}}}
\newcommand{\lljj}{\ensuremath{\ell\ell \text{jj}}}
\newcommand{\eejj}{\ensuremath{\Pe\Pe \text{jj}}}
\newcommand{\mumujj}{\ensuremath{\PGm\PGm \text{jj}}}
\newcommand{\emujj}{\ensuremath{\Pe\PGm \text{jj}}}
\newcommand{\Rllemu}{\ensuremath{R_{\ell\ell/\Pe\PGm}}}

\cmsNoteHeader{EXO-17-011}
\title{Search for a heavy right-handed \PW\ boson and a heavy neutrino in events with two same-flavor leptons and two jets at $\sqrt{s}=13\TeV$}

\date{\today}

\abstract{A search for a heavy right-handed \PW\ boson (\WR) decaying to a heavy right-handed neutrino and a charged lepton in events with two same-flavor leptons (\Pe\ or \PGm) and two jets, is presented. The analysis is based on proton-proton collision data, collected by the CMS Collaboration at the LHC in 2016 and corresponding to an integrated luminosity of 35.9\fbinv. No significant excess above the standard model expectation is seen in the invariant mass distribution of the dilepton plus dijet system. Assuming that couplings are identical to those of the standard model, and that only one heavy neutrino flavor \NR contributes significantly to the \WR decay width, the region in the two-dimensional (\mWR, \mNR) mass plane excluded at 95\% confidence level extends to approximately $\mWR=4.4\TeV$ and covers a large range of right-handed neutrino masses below the \WR boson mass. This analysis provides the most stringent limits on the \WR mass to date.
}

\hypersetup{
pdfauthor={CMS Collaboration},
pdftitle={Search for a heavy right-handed W boson and a heavy neutrino in events with two same-flavor leptons and two jets at sqrt(s)=13 TeV},
pdfsubject={CMS},
pdfkeywords={CMS, physics, heavy neutrino, right-handed bosons}}

\maketitle

\section{Introduction}
\label{sec:intro}
Heavy partners of the standard model (SM) gauge bosons, that are coupled to right-handed fermions, are predicted in left-right (LR) symmetric models~\cite{Pati:1974yy, Mohapatra:1974gc, Senjanovic:1975rk,Keung:1983uu}. These models explain the parity violation observed in weak interactions as the consequence of spontaneous symmetry breaking at a multi-TeV mass scale.
This paper describes a search for such a heavy partner, a heavy right-handed gauge boson \WR, in events with two same-flavor leptons (\Pe\ or \PGm) and two jets. The study was conducted by the CMS Collaboration at the CERN LHC, using proton-proton collision data corresponding to an integrated luminosity of 35.9\fbinv recorded during the 2016 data taking period.

The right-handed bosons are assumed to interact with the SM particles with a coupling strength \gR. This is a free parameter in most LR models, but we assume a strict LR symmetry in our search so that the coupling constant \gR\ is the same as the SM coupling constant \gL. We also assume that the right-handed quark mixing matrix is the same as the Cabibbo--Kobayashi--Maskawa matrix. In addition to the gauge bosons, LR models usually include heavy right-handed neutrinos (\NR)~\cite{Adhya:2000us,Dev:2016qeb}.
The existence of these heavy neutrinos can explain the very small masses of the SM neutrinos as a consequence of the see-saw mechanism~\cite{Mohapatra:1979ia, Das:2017hmg, GellMann:1980vs}.

In this search, we consider the case in which the \WR boson decays to a first- or second-generation charged lepton and a heavy neutrino of the same lepton flavor. The heavy neutrino further decays to another charged lepton of the same flavor and a virtual \WRv. The virtual \WRv\ decays to two light quarks, producing the decay chain
\begin{linenomath}
\begin{equation*}
\WR\to \ell \NR \to \ell\ell \WRv \to \ell\ell \Pq \Paq^\prime, \,\ell = \Pe\,\text{or}\,\PGm.
\end{equation*}
\end{linenomath}
The quarks hadronize into jets that can be observed by the CMS detector.
The lepton flavor is conserved, and there is no charge requirement on the leptons, which can be opposite-sign or same-sign.
The SM processes that have the same final state of two same-flavor leptons and two jets include Drell--Yan production of lepton pairs with additional jets (DY+jets), \ttbar production, \cPqt\PW\ from $t$-channel single top quark production, and diboson production (\PW\PZ, $\PZ\PZ$, \PW\PW) with jets. Contributions due to  events with jets misidentified as leptons are considered, but are found to be negligible. The discriminating variable in this search is the invariant mass \mlljj\ constructed from the two leptons and two jets with the largest transverse momenta.
We search for an excess of events above the SM prediction for different \WR mass hypotheses in windows of \mlljj.

A search for \WR  bosons that was performed by the CMS Collaboration at a center-of-mass energy of $\sqrt{s}=8\TeV$ excluded \WR masses up to approximately 3\TeV at 95\% confidence level (\CL)~\cite{Khachatryan:2014dka}. An excess with a local significance of 2.8$\sigma$ was observed in that search in the electron channel at $\meejj\approx2.1\TeV$. The excess did not appear to be consistent with signal events from the LR symmetric theory.
The search presented in this paper extends this previous search using data collected at $\sqrt{s}=13\TeV$ during 2016.
It does not overlap with other heavy neutrino searches previously carried out by the CMS Collaboration~\cite{Khachatryan:2016jqo,Sirunyan:2017yrk,Sirunyan:2017xnz}.
The ATLAS Collaboration has also carried out similar searches~\cite{ATLAS:2012ak,ATLAS:2013oea,Aad:2015xaa}.

\section{The CMS detector}
\label{sec:apparatus}
The central feature of the CMS apparatus is a superconducting solenoid of 6\unit{m} internal diameter, providing a magnetic field of 3.8\unit{T}. Within the solenoid volume are a silicon pixel and strip tracker, a lead tungstate crystal electromagnetic calorimeter (ECAL), and a brass and scintillator hadron calorimeter (HCAL), each composed of a barrel and two endcap sections.
Forward calorimeters extend the pseudorapidity ($\eta$) coverage provided by the barrel and endcap detectors.
Electrons are measured in the ECAL, while drift tubes, cathode strip chambers, and resistive-plate chambers embedded in the steel flux-return yoke outside the solenoid are used in the identification of muons.
A more detailed description of the CMS detector, together with a definition of the coordinate system used and the relevant kinematic variables, can be found in Ref.~\cite{Chatrchyan:2008zzk}.

\section{Trigger, particle reconstruction, and event selection}
\label{sec:eventsel}
Events of interest are selected online using a two-tiered trigger system~\cite{Khachatryan:2016bia}. The first level is composed of custom hardware processors and uses information from the calorimeters and muon detectors to select events at a rate of around 100\unit{kHz}.
The second level consists of a farm of processors running a version of the full event reconstruction software optimized for fast processing, and reduces the event rate to less than 1\unit{kHz} before data storage.

The leptons in the final state carry a large fraction of the rest energy of the \WR. Thus, a trigger with a high momentum requirement on the lepton is highly efficient for our signal.
For events with electrons, we use an unprescaled double-electron trigger. This trigger requires a minimum transverse momentum (\PT) of 33\GeV and an ECAL energy deposit with a pixel hit on an associated track.
For the muon channel, and for an auxiliary measurement that is used to estimate the \ttbar background, we use unprescaled single-muon triggers that have no isolation requirement and a $\PT>50\GeV$ requirement applied to the muon.

Global event reconstruction is performed using the particle-flow algorithm~\cite{CMS-PRF-14-001}, which reconstructs and identifies each individual particle with an optimized combination of all subdetector information.
At least one reconstructed vertex is required.
For events with multiple collision vertices from additional collisions in the same or adjacent bunch crossings (pileup interactions), the reconstructed vertex with the largest value of summed $\PT^2$ in the event, where the sum extends over all charged tracks associated with the vertex, is taken to be the primary $\Pp\Pp$ interaction vertex (PV).

Electron candidates are identified by the association of a charged-particle track from the PV, with energy deposits clusters (superclusters) in the ECAL. The association takes into account energy deposits both from the electron and from bremsstrahlung photons produced during its passage through the inner detector.
The electron momentum is estimated by combining the energy measurement in the ECAL with the momentum measurement in the tracker.
The experimental mass resolution for barrel-barrel (barrel-endcap) dielectron pairs with a mass of 1\TeV is 1.0\,(1.5)\%~\cite{Khachatryan:2015hwa}.
To correct for observed discrepancies in energy scale and resolution between data and simulation, the measured electron energy is adjusted  by a multiplicative factor that depends on $\eta$ and \RNINE, where \RNINE\ is the ratio of the energy in a 3$\times$3 matrix of ECAL crystals, centered on the crystal with the largest energy, to the full energy collected by a supercluster.
In addition, the electron energy in simulated events is smeared by 1--3\% using a Gaussian expression that varies as a function of $\eta$ and \RNINE ~\cite{Khachatryan:2015hwa}.
Differences in electron identification (ID) efficiency between data and simulation were taken into account by applying a scale factor (SF) of $0.972 \pm 0.006$\,(stat+syst) in the barrel and $0.983 \pm 0.007$ (stat+syst) in the endcaps.

Muons are reconstructed from tracker and muon chamber information.
Each muon is required to have at least one hit in the pixel detector, at least six tracker layer hits, and segments in two or more muon detector stations.
Muons are measured in the range $\abs{\eta} < 2.4$.
The \pt resolution in the barrel is better than 10\% for muons with \pt up to 1\TeV~\cite{Chatrchyan:2012xi}.
The muon momentum resolution in data is well described in simulated events, with its uncertainty provided by a smearing of 1\% in the barrel and 2\% in the endcaps.
The muon curvature distributions in data and simulation are compared for different ranges of $\eta$ and azimuthal angle ($\phi$, in radians), resulting in the assignment of a momentum scale uncertainty of 3\% in the barrel and up to 9\% in the endcaps.
To account for differences in the reconstruction and identification efficiencies between data and simulation, $\eta$-dependent SFs in the range 0.95--0.99 are applied to simulated events~\cite{Chatrchyan:2012xi}.
Systematic uncertainties related to the dependence of the SFs on momentum are neglected, since they have an impact on the results of less than 1\%.

Charged hadrons are identified by matching tracks to one or more calorimeter clusters, and by the absence of signal in the muon detectors.
The energies of charged hadrons are determined from combinations of the track momenta and the corresponding ECAL and HCAL energies,
corrected for zero-suppression effects and for the response function of the calorimeters to hadronic showers.

Neutral hadrons are identified as ECAL and HCAL energy clusters that are not matched to charged particle trajectories. The energies of neutral hadrons are obtained from the corresponding corrected ECAL and HCAL energies.

For each event, hadronic jets are clustered from reconstructed particles with the anti-\kt algorithm, operated with a size parameter $R$ of 0.4, where $R\equiv\sqrt{\smash[b]{(\Delta\eta)^2+(\Delta\phi)^2}}$~\cite{Cacciari:2008gp, Cacciari:2011ma}.
Charged hadrons that originate from pileup interactions
are removed from the list of reconstructed particles using the charged-hadron pileup subtraction algorithm~\cite{CMS-PRF-14-001}.
The contributions of neutral particles that originate from pileup interactions to the calorimeter energies are removed by applying a residual average area-based correction~\cite{Cacciari:2007fd}.
The jet momentum is defined as the vector sum of all particle momenta associated with the jet, and is found to be within 5
to 10\% of the true momentum in simulated events over the whole \pt spectrum and detector acceptance.
Jet energy corrections are derived from the simulation, and are confirmed with in-situ measurements of the energy balance in dijet, multijet, photon+jet, and leptonically decaying \cPZ+jet events~\cite{Khachatryan:2016kdb}.
Jet identification algorithms~\cite{Khachatryan:2016kdb} also remove contributions to jets from calorimeter noise and beam halo.

To reconstruct \WR candidates, we select the two leptons with the largest \pt and the two jets with the largest \pt.
The leading (subleading) leptons are required to have $\PT > 60\,(53) \GeV$ and to be within the detector acceptance ($\abs{\eta}<2.4$).
Electrons are rejected if the supercluster lies in the range $1.444 < \abs{\eta} < 1.566$, which corresponds to the transition region between the barrel and endcap sections of the ECAL, where the performance is degraded.
To suppress muons originating from hadron decays or pion punch-through in jets, we remove muons for which the sum of the \pt of additional tracks that originate from the PV and that are inside a cone of $R < 0.3$ around the muon is more than 10\% of the muon \PT.
We also require electrons to be isolated, \ie, the sum of the \pt of all tracks inside a cone of $R < 0.3$ centered on the electron candidate,  not associated with the electron and originating from the PV must be below 5\GeV.
We use dedicated identification algorithms, optimized for the selection of high-momentum leptons~\cite{Chatrchyan:2012xi,Khachatryan:2015hwa}.
The two jet candidates must each have $\pt>40\GeV$ and be within $\abs{\eta}< 2.4$.
To avoid having reconstructed leptons overlap jets, we impose a $\Delta R> 0.4$ requirement between all jets and leptons.

\section{Signal model}
\label{sec:samples}
We use several auxiliary data samples to estimate signal and background contributions to our search as well as to validate our event selection. We use Monte Carlo (MC) simulation in the calculation of the signal efficiency and in the estimation of some of the SM backgrounds. In these simulations, the response of the CMS detector is modeled using the \GEANTfour package~\cite{Agostinelli:2002hh}. Pileup contributions are modeled by superimposing simulated inelastic proton-proton interactions onto the primary hard scattering. The simulated distribution of the number of pileup events is matched to that observed in the data.

For estimating the acceptance and efficiency for detecting \WR bosons, simulated signal samples of \eejj\ and \mumujj\ final states are generated assuming $\mNR = 1/2 \mWR$, using the \PYTHIA 8.212 program~\cite{Sjostrand:2014zea} with the NNPDF2.3~\cite{Ball:2012cx} parton distribution functions (PDFs).
Simulated signal samples with $\mNR \neq 1/2 \mWR$, needed to estimate the 2D limits described in Section~\ref{sec:results}, are also generated using \PYTHIA 8.212.

We focus our search on a region of phase-space where the signal is expected to appear. This signal region applies to events with two leptons with the same flavor and two jets. The invariant mass of the dilepton system must be above 200\GeV, to avoid contamination from resonant \PZ boson production. The \mlljj\ must be greater than 600\GeV to ensure that all the kinematic requirements on the candidates are fully efficient.
There is no charge requirement on the leptons, to ensure sensitivity to a wider class of models.

Using the selection requirements described above, the product of the acceptance and efficiency for \WR decays to the \lljj\ final state, increases from 30\% at $\mWR = 1000\GeV$ to 57\% for $\mWR >3000\GeV$ in the electron channel, and similarly from 40 to 75\% in the muon channel. For both channels, the signal efficiency reaches a plateau at $\mWR = 3000\GeV$. The efficiency for electron events is lower than the muon event efficiency because of differences between the selection requirements, and because of the omission of the transition regions between the ECAL barrel and endcaps in the case of electrons.

\section{Background estimation}
\label{sec:backgrounds}
Standard model processes that produce events with the same final-state particles as the signal model include DY production of lepton pairs with additional jets in the final state, and \ttbar and diboson production. The DY+jets and \ttbar production are irreducible background processes that comprise most of the background events in the signal region.
The contribution from diboson backgrounds is suppressed by the dilepton mass requirement ($\mll>200\GeV$). We also consider backgrounds for which candidate misidentification leads to events with two leptons and two jets in the final state. These backgrounds include \PW\ boson production with additional jets, $t$-channel single top quark events with additional jets, and QCD multijet events. These reducible backgrounds do not significantly contaminate our signal region.
The diboson backgrounds constitute $\sim$1.5\% of the total background in the signal region, the \PW+jets $\sim$0.5\%, the single top quark events $\sim$5\%, and the QCD events $\sim$0.1\%.

The MC samples used to estimate the background processes are simulated with several MC event generators.
The DY+jets and the \ttbar samples are generated with \MGvATNLO 2.3.3~\cite{Alwall:2014hca} at next-to-leading order (NLO) using the NLO NNPDF3.0~\cite{Ball:2014uwa} PDF set.
Diboson (\PW\PW, \PW\PZ, and $\PZ\PZ$) samples are generated at leading order (LO) using \PYTHIA 8.212 along with the LO NNPDF2.3~\cite{Ball:2012cx} PDFs, while \PW+jets events are generated with \MGvATNLO 2.3.3~\cite{Alwall:2014hca} at leading order (LO) and single top quark events are produced in the \cPqt\PW\ channel with \POWHEG v1.0~\cite{Nason:2004rx,Frixione:2007vw,Alioli:2010xd,Re:2010bp}.
The more precise NLO calculations are used to normalize the SM simulated samples of diboson, \PW+jets and single top quark events to NLO accuracy.
The NNPDF3.0 PDFs are used for samples generated at NLO.
For all samples, \PYTHIA 8.212 is used for parton showering, fragmentation and hadronization with the underlying event tune \textsc{cuetp8m1} \cite{Khachatryan:2015pea}.
The DY+jets samples have one parton at the matrix element level, and additional parton showering is modeled in \PYTHIA.
The potential double counting of partons generated using \PYTHIA with those using \MGvATNLO is minimized using the MLM~\cite{Alwall:2007fs} (FXFX~\cite{Frederix:2012ps}) matching scheme in the LO (NLO) samples.

We define different regions of phase-space (control regions) to estimate the contributions of the different SM backgrounds.
To study the background contribution from DY+jets events we use a sample defined by the presence of two same-flavor, opposite-charge electrons or muons and two jets. The invariant mass of the dilepton system must satisfy $\mll<200\GeV$. We call this the ``low dilepton mass control region''.
The ``flavor control region'', used to study the \ttbar background contribution, corresponds to an event sample composed of one electron, one muon, and two jets. For this region the invariant mass of the dilepton system must satisfy $\mll>200\GeV$, while the \mlljj\ is required to be above 600\GeV.

\subsection{Drell--Yan background}
Monte Carlo simulation is used to estimate the background from high mass DY lepton pair production in association with additional jets, since no high purity control region has been identified having the same kinematic characteristics as the signal region.
The normalization of DY+jets background in simulation is adjusted to match the event counts in data using a SF calculated as the ratio of data and simulation events under the \PZ resonance in the range $80<\mll<100\GeV$.
This SF corrects for residual mismodeling between data and simulation, and includes the signal region requirements on the jets. The measured SF is $1.07\pm 0.01\stat$ in both electron and muon channels.

\begin{figure}[tbp]
  \centering
  \includegraphics[width=0.45\textwidth]{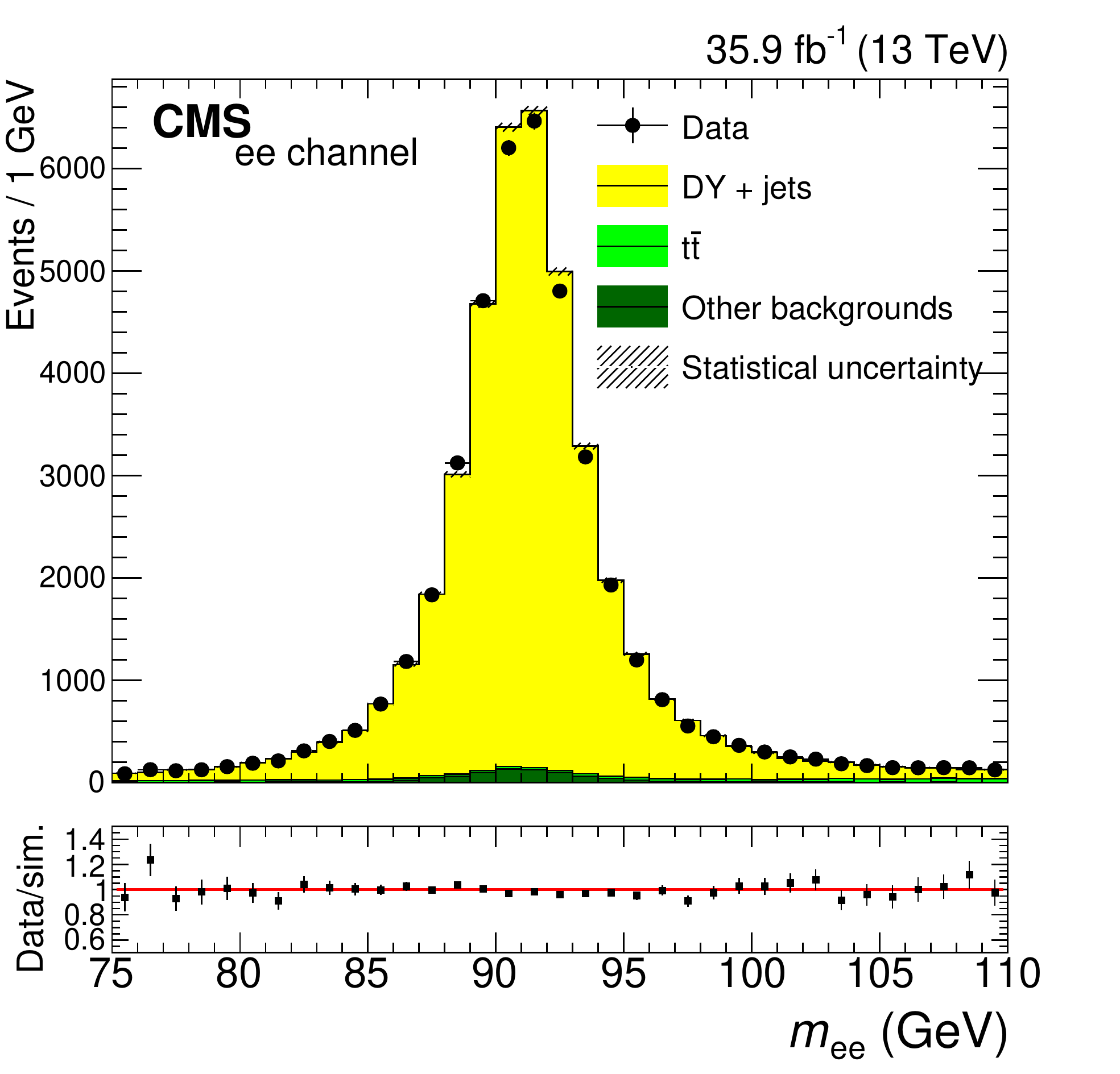}
  \includegraphics[width=0.45\textwidth]{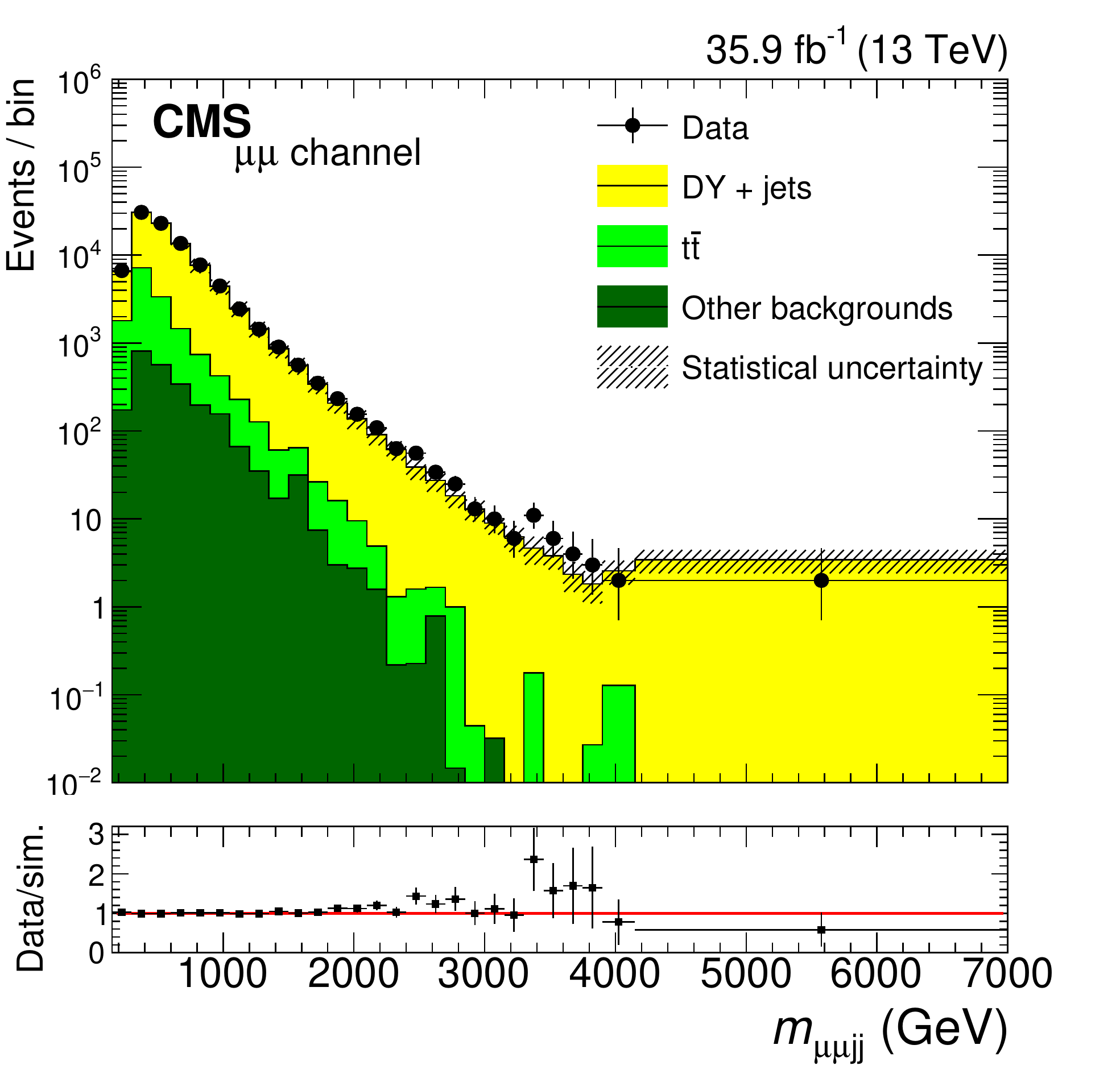} \\
  \includegraphics[width=0.45\textwidth]{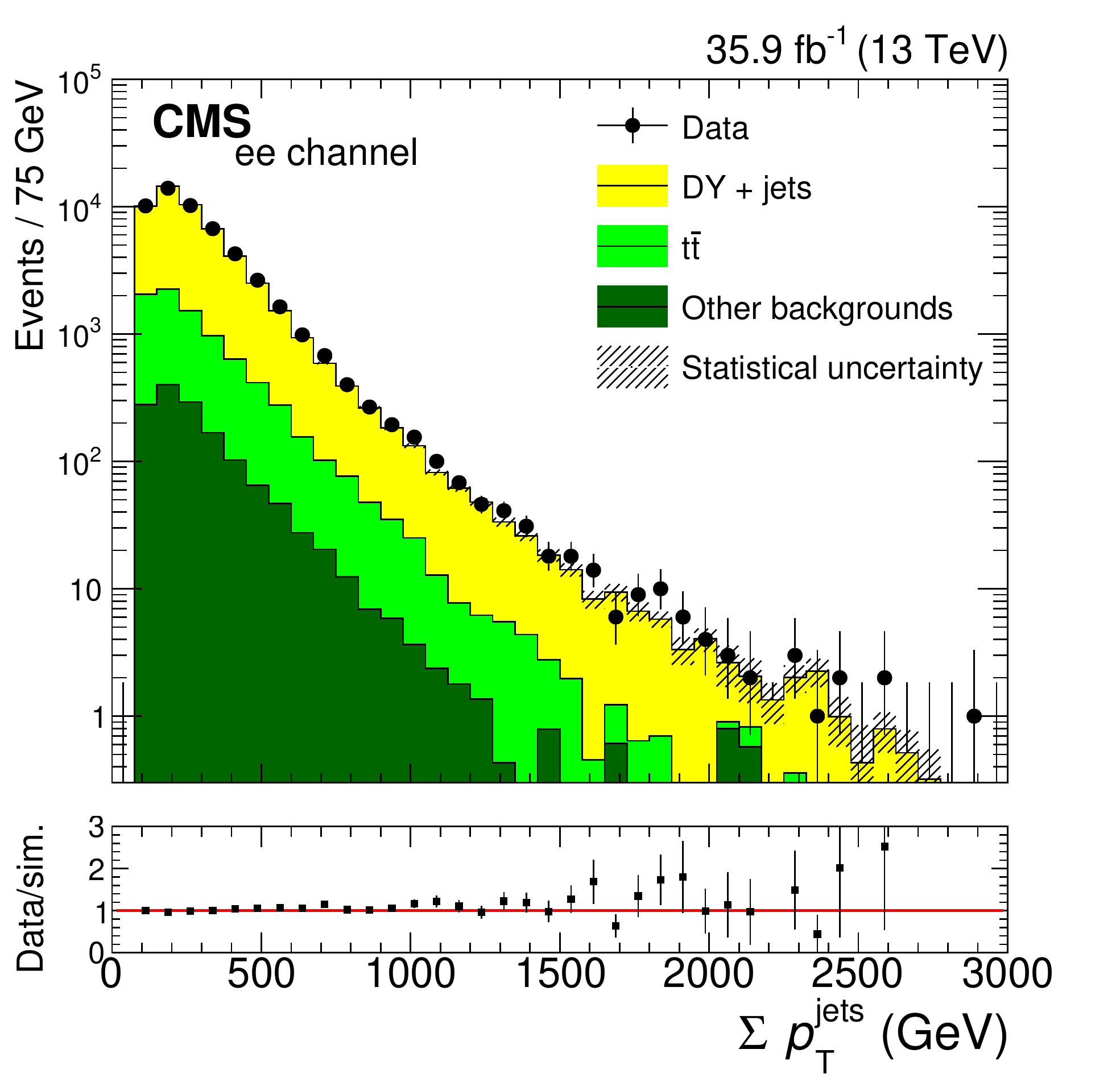}
  \includegraphics[width=0.45\textwidth]{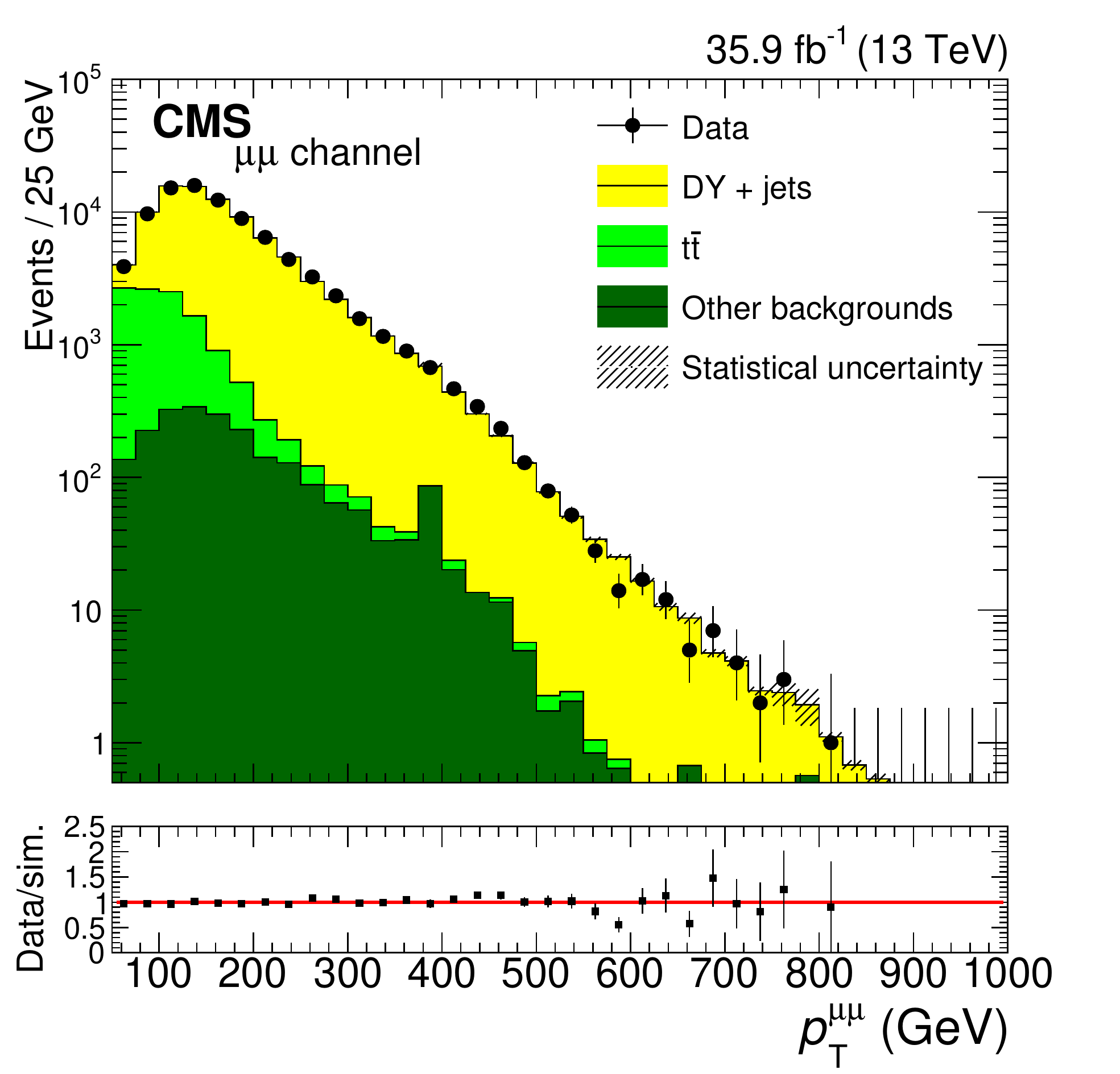}
  \caption{Kinematic distributions for events in the low dilepton mass control region with the DY SF applied.
  The dilepton mass (upper left) and the scalar sum of all jet transverse momenta (lower left) are shown for the $\Pe\Pe$ DY plus two jets selection. 
  The \mlljj\ (upper right) and the dilepton transverse momentum (lower right) are shown for the $\PGm\PGm$ DY plus two jets selection.
  The uncertainty bands on the simulated background histograms include only statistical uncertainties. The uncertainty bars in the ratio plots represent combined statistical uncertainties of data and simulation.}
  \label{fig:lowdilepSB}
\end{figure}

We compare between data and MC all the kinematic distributions of the low dilepton mass control region for the \Pe\Pe\, and $\PGm\PGm$ channels, respectively.
The agreement in this control region is especially important since we derive the estimate for the shape of the DY+jets background directly from simulation.
The distributions of some kinematic quantities in the low dilepton mass control region with the SF already applied are shown in Fig.~\ref{fig:lowdilepSB} for both electron and muon channels.
In these plots, all expected SM backgrounds, except for DY+jets and \ttbar, are labelled as \textit{Other backgrounds}. Good agreement is observed in the shapes of the kinematic distributions in both cases.

\begin{figure}[tbp]
  \centering
  \includegraphics[width=0.45\textwidth]{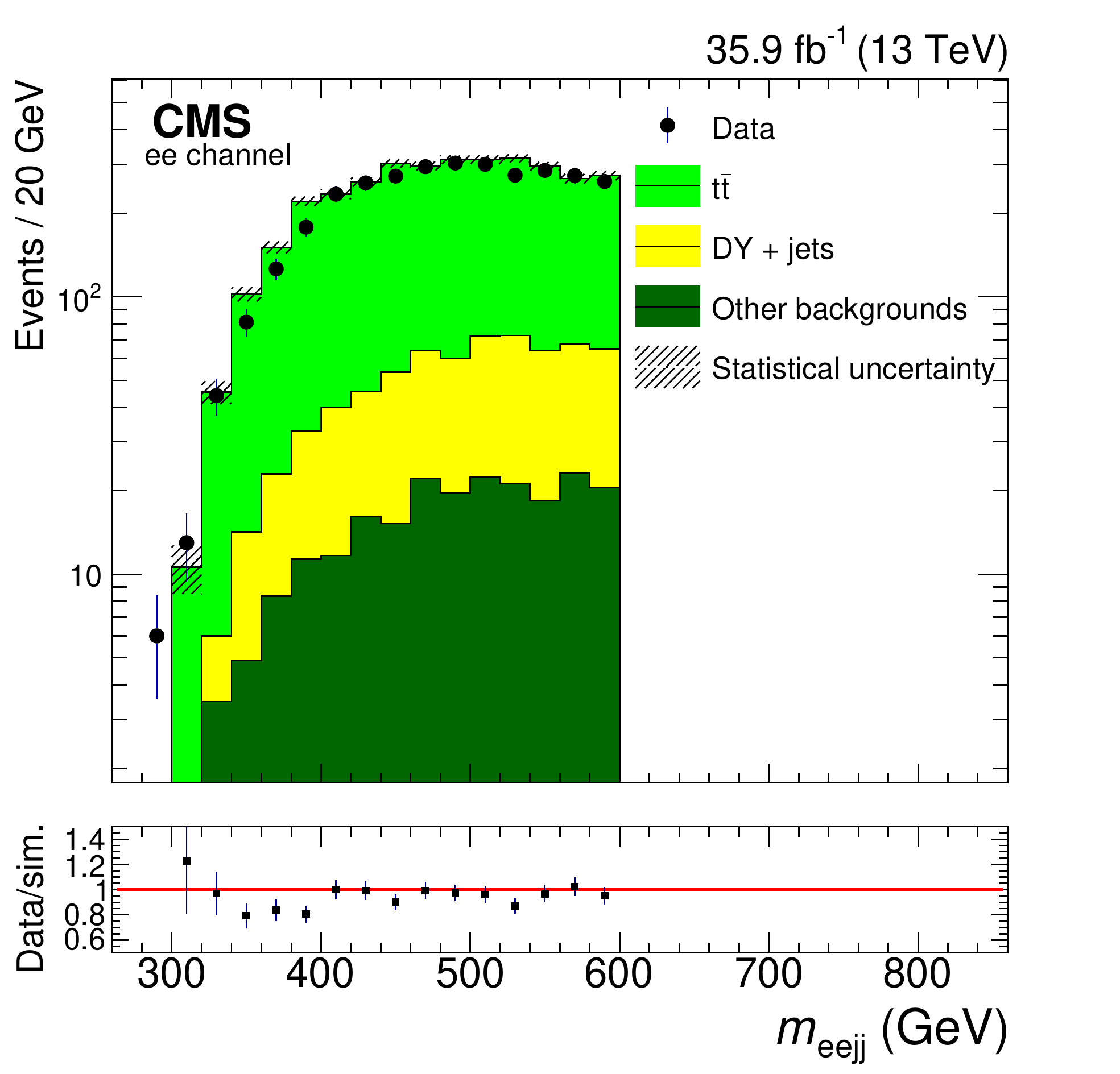}
  \includegraphics[width=0.45\textwidth]{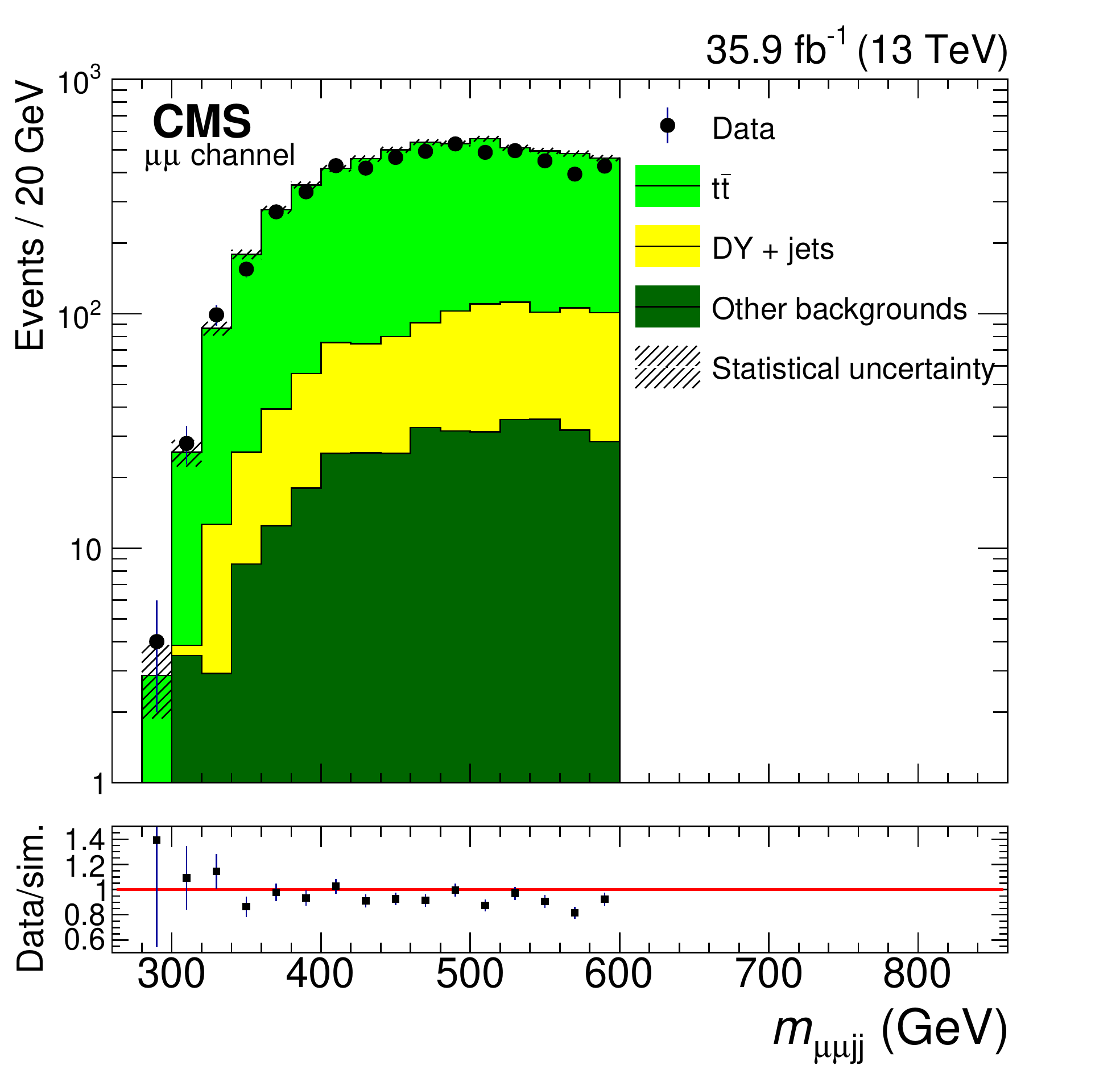}
  \caption{The \mlljj\ distribution in the low \mlljj\ control region with the DY SF applied for the electron (left) and muon (right) channel.
  The uncertainty bands on the simulated background histograms include only statistical uncertainties. The uncertainty bars in the ratio plots represent combined statistical uncertainties of data and simulation.}
  \label{fig:lowMlljj}
\end{figure}

To verify that the SF measured for DY+jets below the \PZ boson peak is valid also at higher dilepton masses, we use a dedicated control region, referred to as the ``low \mlljj\ control region'', which is defined by the signal region selections, except for an inverted $\mlljj<600\GeV$ requirement.
In this control region, we check for agreement between data and simulation in events with high dilepton mass.
The \mlljj\ distributions with the DY SF applied are shown in Fig.~\ref{fig:lowMlljj}.

\subsection{\texorpdfstring{\ttbar}{ttbar} background}
The \ttbar background contribution is estimated directly from data in the flavor control region defined above, which has the same kinematic characteristics as the \ttbar events in the signal region.
For this estimate, we use the events in the flavor control region, assuming that there is no contamination from signal events.
This assumption, which corresponds to an imposition of the conservation of individual lepton flavor on our signal models, is valid since, at leading order, the decay of a \WR boson cannot yield events with an \emujj\ final state.

To calculate the number of events from \ttbar production in the \eejj\ and the \mumujj\ signal regions, we use simulated \ttbar events to determine transfer factors \Rllemu\ ($\ell\ell=\Pe\Pe$ or $\PGm\PGm$) between the \emujj\ control region and the signal region.
These factors are evaluated from the ratio of the number of simulated \ttbar events in the distributions of \meejj\ or $m_{\mumujj}$ in the signal region to the number of events in the distribution of $m_{\emujj}$ in the flavor control region. The number of events in the signal region is then given by:
\begin{equation}
  N_{\ttbar}(\text{signal region})=N_{\ttbar}(\text{flavor control region})\, \Rllemu.
\end{equation}
Using the transfer factor, we can account for the difference in the efficiency and acceptance between electrons and muons in these final states.
\begin{table}[tb]
\topcaption{Transfer factors applied to the number of events in the flavor control region to estimate the number of \ttbar events in the \eejj\ and \mumujj\ signal regions.}
\label{tab:ttbarSF}
\centering
\begin{tabular}{lccc}
Channel          & Transfer factor & Stat. uncertainty & Syst. uncertainty \\
\hline
 $\emujj \to \eejj$ & 0.42 & 0.01 & 0.07 \\
 $\emujj \to \mumujj$ & 0.72 & 0.02 & 0.14 \\
\end{tabular}
\end{table}
The values of the transfer factors obtained are given in Table~\ref{tab:ttbarSF}.
The \Rllemu\ as a function of the \mlljj\ distribution is fit to a constant. A systematic uncertainty is assigned by fitting the transfer factor to a linear function and taking the difference between the values of this function at the high and low \mlljj.
\begin{figure}[tbp]
  \centering
  \includegraphics[width=0.45\textwidth]{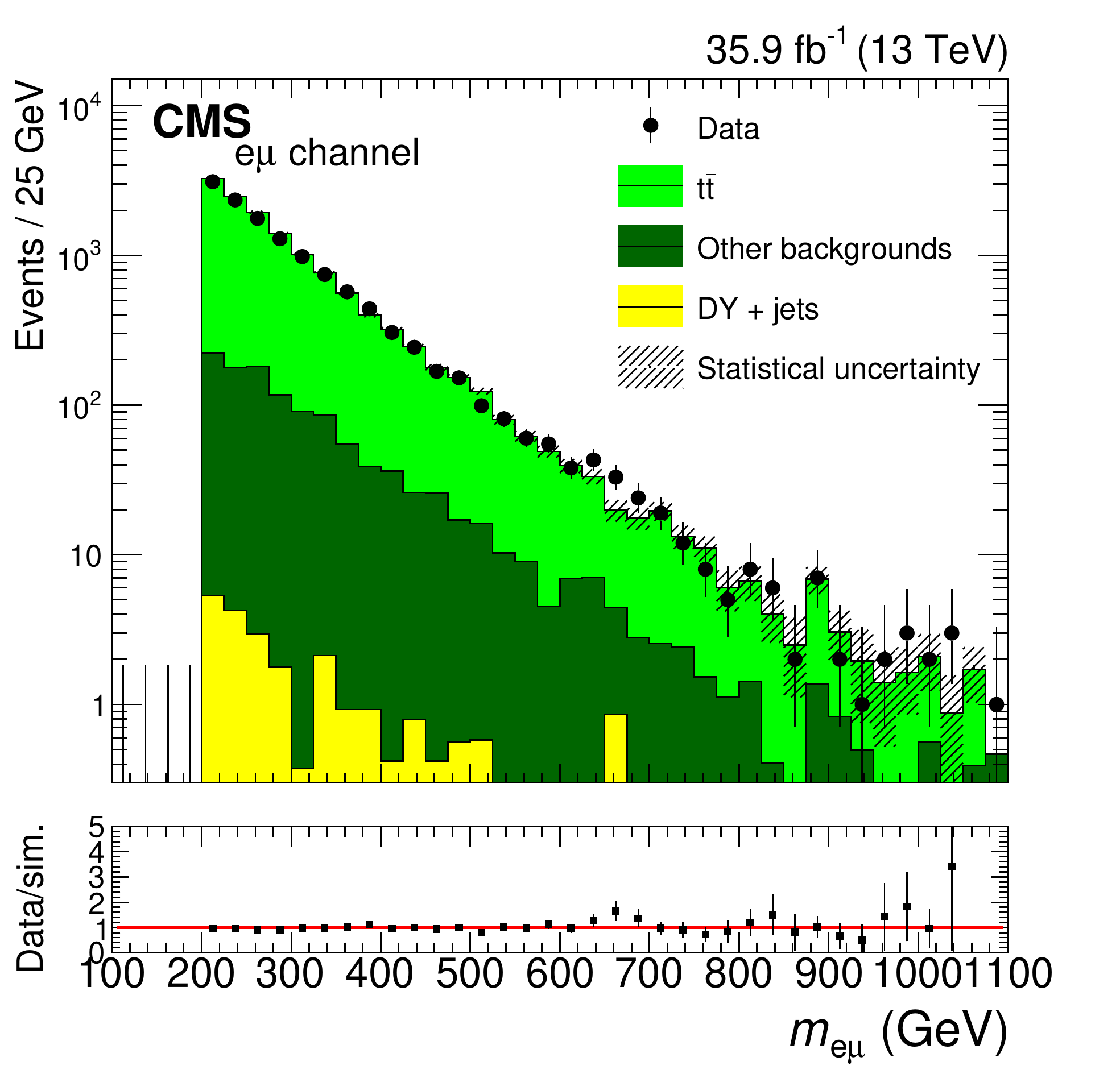}
  \includegraphics[width=0.45\textwidth]{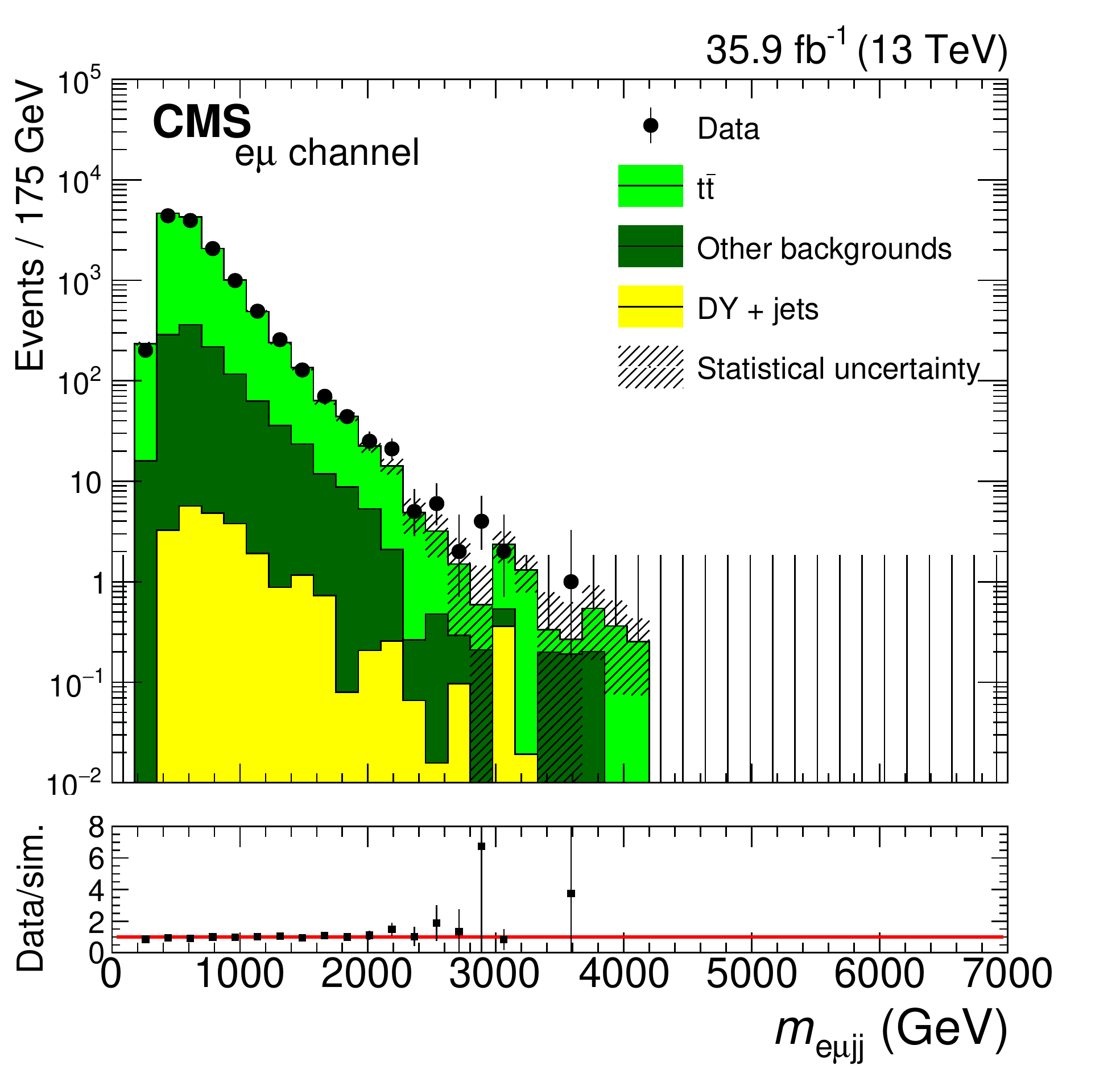}\\
  \includegraphics[width=0.45\textwidth]{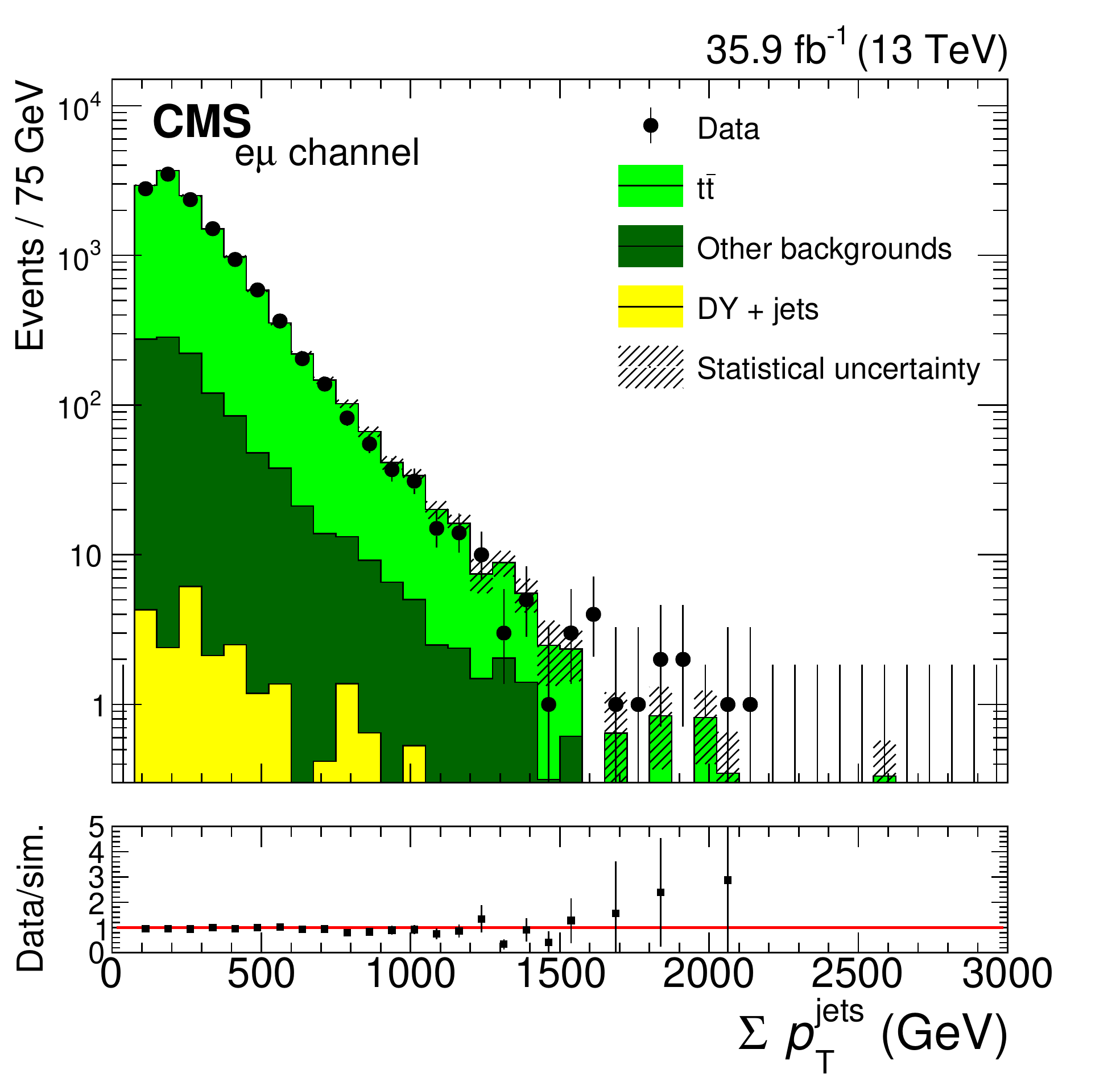}
  \includegraphics[width=0.45\textwidth]{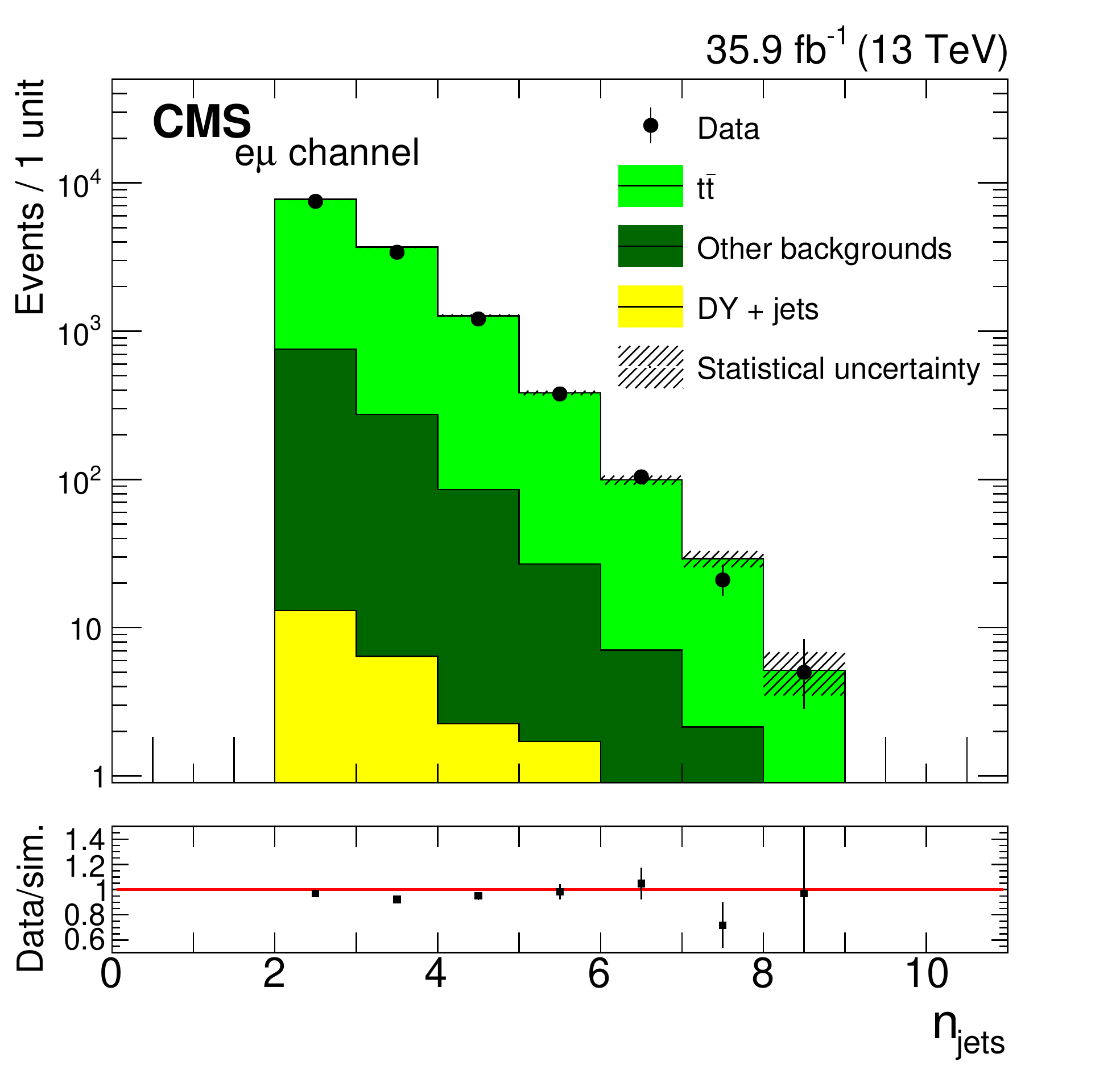}
  \caption{ Kinematic distributions for events in the flavor control region  with the DY SF applied.
  The dilepton mass (upper left), the \mlljj\ (upper right), the scalar sum of all jet transverse momenta (lower left), and the number of jets (lower right) are shown.
  The uncertainty bands on the simulated background histograms include only statistical uncertainties. The uncertainty bars in the ratio plots represent combined statistical uncertainties of data and simulation.}
  \label{fig:flavorSB}
\end{figure}
Figure~\ref{fig:flavorSB} shows a comparison between simulated events and data for several kinematic variables in the flavor control region.

The \ttbar background contribution in the signal region is estimated without the direct use of simulated events.
However, the agreement between simulation and data in the flavor control region suggests that other modeling using simulation, such as the signal acceptance, is reliable.

\section{Results}
\label{sec:results}
The strategy followed in this analysis is to search for deviations from the shape of the \mlljj\ distribution expected in the standard model. This distribution extends over a range of several TeV. While the LR symmetric models motivate the choice of the \lljj\ final state, we do not impose requirements on the signal shape specific to these models, in order to maintain sensitivity to other models.
The strategy to search for an excess of events in a wide mass range is effective in analyzing the data without exploiting other characteristics of the benchmark signal model and reduces the effect of the uncertainties in the shapes of the backgrounds, especially in the high-\mlljj\ region.
The expected number of signal and background events is estimated by counting the events falling in a particular \mlljj\ window.
The upper and lower limits of the mass window are chosen as a function of \mWR\ to obtain the most stringent expected cross section upper limits.
Optimizing with respect to signal significance instead results in comparable mass windows.
The width of the mass window for the electron final state varies from 130\GeV at low masses ($\mWR\simeq800\GeV$) to 3100\GeV at high masses ($\mWR\simeq6000\GeV$).
For muons, the mass window varies more, and becomes as large as 3800\GeV.
The upper and lower bounds are fitted as functions of \mWR\ to third degree polynomials to reduce the effect of statistical fluctuations in the optimization procedure.

The probability of the observed number of events being produced by a combination of background and signal with a cross section $\sigma$ is calculated using a Bayesian approach with flat signal prior and a fit model with nuisance parameters introduced to address the uncertainties,  with log-normal priors.
The exclusion limit on the cross section $\sigma$ is defined as the upper bound of the one-sided 95\% credibility interval determined from the posterior likelihood distribution for the signal cross section.
This procedure is repeated for each mass hypothesis.

In order to take into account the statistical and systematic uncertainties, pseudo-experiments are performed, varying the expected number of events from signal and background according to the uncertainties as described below.
The median of the distribution of the excluded cross section produced by pseudo-experiments and the intervals containing 68 and 95\% of the pseudo-experiments are then quoted in the expected limits and their uncertainties.

The sources of systematic uncertainty considered in this analysis are the uncertainty in the integrated luminosity determination~\cite{cms_lumi}, the normalization uncertainty in the \ttbar background, the uncertainties due to proton PDFs, and factorization and renormalization scales for the DY+jets background and the signal, and the systematic effects related to candidate reconstruction.
This last set of uncertainties, affecting the shape of the \mlljj\ distribution, include uncertainties in the jet and lepton energy scales and resolutions, and in the lepton reconstruction, trigger, isolation, and identification SFs.

In order to propagate the uncertainties in candidate reconstruction, a large number of pseudo-experiments are performed, varying all the uncertainty sources at the same time in an uncorrelated fashion, each according to a Gaussian distribution with mean equal to the nominal value and width equal to the uncertainty of the single source. The variations are performed before the event selection, so each pseudo-experiment is processed using the full analysis chain. The expected number of events for signal and background in a mass window is evaluated for each pseudo-experiment.
The values used to extract the limit are given by the mean of the pseudo-experiment distribution, and their standard deviation is the propagated uncertainty.
The uncertainties in the candidate reconstruction are then implemented as nuisance parameters with log-normal priors in the limit evaluation.
The effects of these uncertainties on the signal and background yields are listed in Table~\ref{tab:uncTable}.

\begin{table}[htp]
  \topcaption{Effect of systematic uncertainties in candidate reconstruction efficiencies, energy scale and resolutions on the signal and background yields.
  The Signal column shows the range of uncertainties computed at each of the \WR mass points. The Background column indicates the range of the uncertainties for the backgrounds.}
  \label{tab:uncTable}
  \centering
  \begin{tabular}{l r@{--}l c r@{--}l}
  Uncertainty & \multicolumn{2}{c}{Signal (\%)} & \multicolumn{3}{c}{Background (\%)} \\
    \hline
    Jet energy resolution & ~3.2 & 26 & & 0.90 & 25 \\
    Jet energy scale & ~0.20 & 29 & & 4.8 & 27 \\
    Electron energy resolution & ~3.7 & 4.8 & & 2.7 & 4.5 \\
    Electron energy scale & ~3.7 & 6.4 & & 4.9 & 5.9 \\
    Electron reco/trigger/ID & ~8.7 & 11 & & 6.1 & 10 \\
    Muon energy resolution & ~4.7 & 10 & & 6.9 & 12 \\
    Muon energy scale & ~4.7 & 10 & & 6.2 & 12 \\
    Muon trigger/ID/iso & ~2.3 & 4.7 & & 1.9 & 5.2 \\
  \end{tabular}
\end{table}

The uncertainty in the integrated luminosity affects only the normalization of the \mlljj\ distributions, as does the uncertainty in the \ttbar extrapolation SF given by the sum in quadrature of its statistical and systematic uncertainties, evaluated as described in Section~\ref{sec:backgrounds}.

The uncertainties in the estimation of the DY+jets background are implemented as a function of \mlljj\ following the PDF4LHC prescription~\cite{Butterworth:2015oua}, and affect both shape and normalization of the \mlljj\ distributions.
Table~\ref{tab:catAuncertainties} lists the range of values of these uncertainties, which are included in the evaluation of the limits as nuisance parameters with log-normal priors.

\begin{table}[tbp]
  \topcaption{Uncertainties affecting the \mlljj\ distribution shape and normalization. The uncertainties in the \ttbar SFs affect the \ttbar background, the uncertainties in the DY PDF and the DY factorization and renormalization scales affect the DY+jets background, and the uncertainty in the integrated luminosity affects both signal and backgrounds.}
  \label{tab:catAuncertainties}
  \centering
    \begin{tabular}{l c}
    Uncertainty                    & \multicolumn{1}{c}{Magnitude (\%)}  \\
      \hline
      \ttbar extrapolation $\Pe\Pe$/$\Pe\PGm$ SF          & 17   (stat+syst) \\
      \ttbar extrapolation $\PGm\PGm$/$\Pe\PGm$ SF        & 20   (stat+syst) \\
      DY $\Pe\Pe$ PDF                     & \,\,15--70  (syst) \\
      DY $\Pe\Pe$ renormalization/factorization & 5.0--40  (syst) \\
      DY $\PGm\PGm$ PDF                     & \,\,10--70  (syst) \\
      DY $\PGm\PGm$ renormalization/factorization & \,\,10--50  (syst) \\
      Integrated luminosity                    & 2.5   (stat+syst) \\
    \end{tabular}
  \end{table}

Concerning uncertainties in the signal arising from the PDF and scale uncertainties, only the effect on the \WR signal acceptance is considered in the expected limit calculation. The effect is implemented as a function of \mlljj\ as for the DY+jets background.

All of the uncertainties that affect the shape of the \mlljj\ distribution also affect the number of events in specific mass ranges and effectively become normalization uncertainties.

To include the statistical uncertainties for each process in the evaluation of the limits, Gamma distributions are used~\cite{CMS-NOTE-2011-005}.
In the limit estimation, pseudo-experiments are generated based on the expected number of events, sampled according to a Gamma distribution and multiplied by the log-normal distributions of the systematics uncertainties.

\begin{figure}[tbp]
  \centering
  \includegraphics[width=0.47\textwidth]{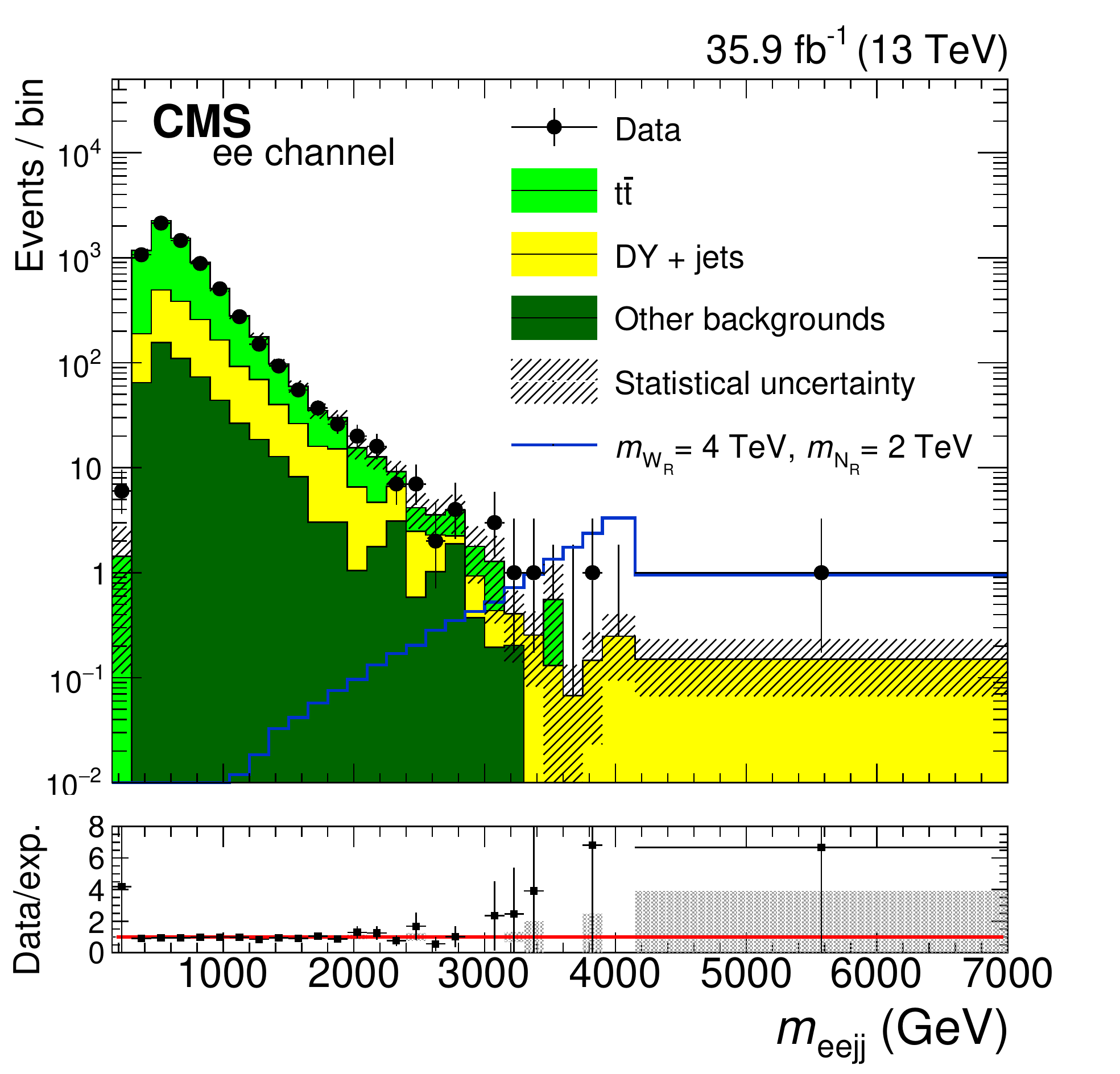}
  \includegraphics[width=0.47\textwidth]{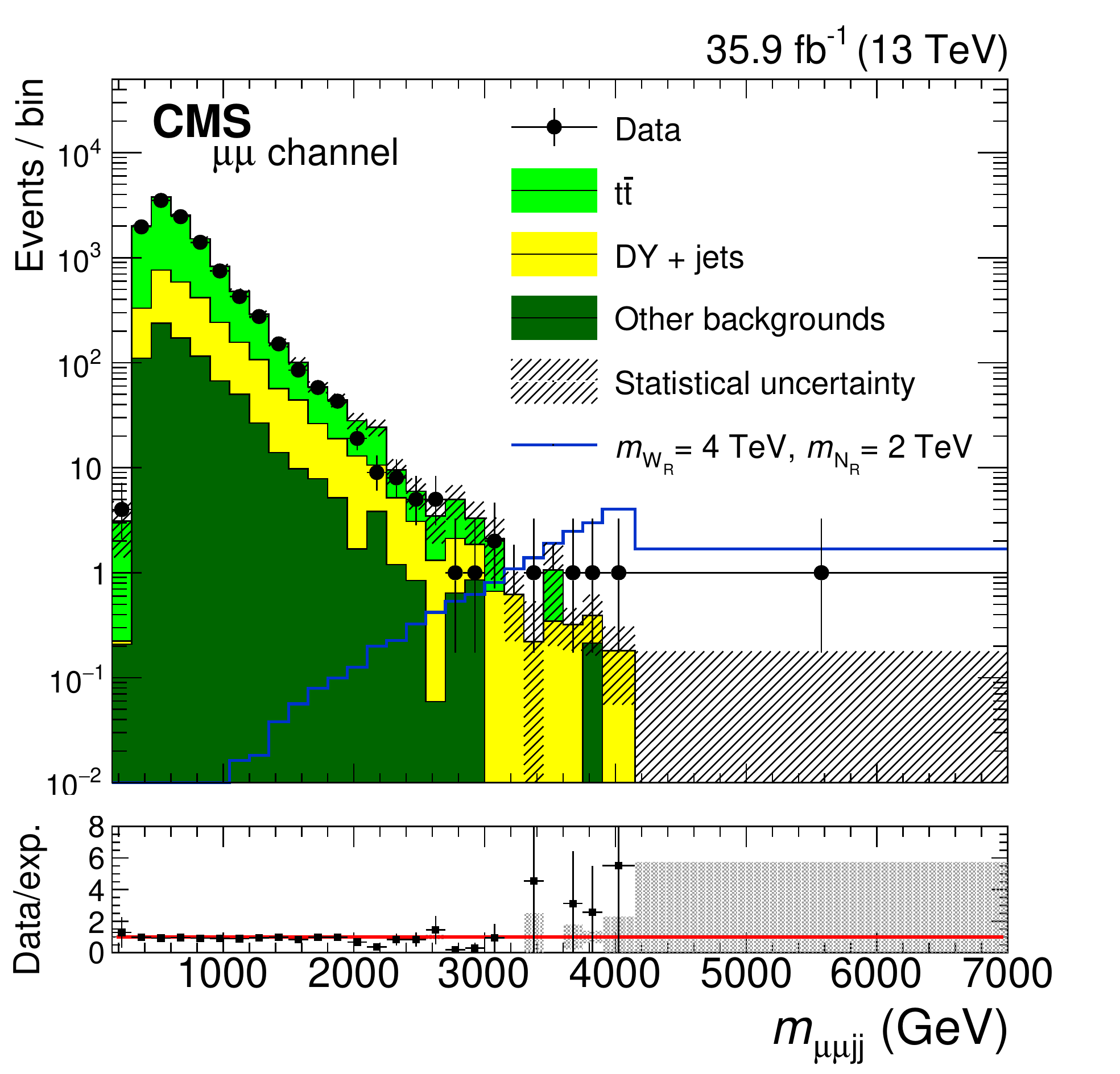}
  \caption{The \mlljj\ distribution in the signal region for the electron (left) and muon (right) channel. The uncertainty bands on the simulated background histograms include only statistical uncertainties. The uncertainty bars in the ratio plots represent combined statistical uncertainties of data and simulation. The gray error band around unity represents the systematic uncertainty on the simulation.}
  \label{fig:signalRegion}
\end{figure}

In Table~\ref{tab:catBuncertainties}, the expected number of events, including the statistical and systematic uncertainties, for the \WR signal, the DY+jets and \ttbar background events, and the total of additional smaller background sources are reported, together with the observed number of events, for several representative \WR mass points.
The signal normalization is obtained for the assumptions $\mNR=1/2\mWR$ and $\gR=\gL$.
In Fig.~\ref{fig:signalRegion}, we present the observed \mlljj\ distribution in the signal region and compare it to the expected backgrounds and the signal shape for $\mWR=4\TeV$.
No significant deviations are seen in the data with respect to expectation.

\begin{table}[tbph]
  \topcaption{Number of expected events for signal, DY+jets, \ttbar, Other, and All backgrounds, as well as the observed number of events in different \WR mass windows.
  All uncertainties are included in the expected number of events. In each table cell, the entry is of the form (mean $\pm$ stat $\pm$ syst).}
  \label{tab:catBuncertainties}

\resizebox{\textwidth}{!}{
\begin{tabular}{c c@{ $\pm$ }c@{ $\pm$ }c c@{ }c@{ $\pm$ }c c@{ }c@{ $\pm$ }c c@{ }c@{ $\pm$ }c c@{ }c@{ $\pm$ }c c}
 $m_\WR$ [mass window] ({\GeVns}) & \multicolumn{3}{c}{Signal} & \multicolumn{3}{c}{DY+jets} & \multicolumn{3}{c}{\ttbar} & \multicolumn{3}{c}{Other} & \multicolumn{3}{c}{All backgrounds} & Data  \\
\hline
\noalign{\vskip 2mm}
\multicolumn{17}{c}{Electron channel}  \\
[\cmsTabSkip]
2200 [1960--2810] & 474.0 & 3.7 & 44.7 & 15.7 & $^{+5.1} _{-3.9}$ & 3.0 & 23.6 & $^{+5.9} _{-4.8}$ & 2.8 & 9.1 & $^{+4.1} _{-2.9}$ & 2.3 & 48.3 & $^{+8.8} _{-6.9}$ & 4.8 & 56 \\
[\cmsTabSkip]
2800 [2530--3840] & 114.1 & 0.9 & 10.6 & 4.1 & $^{+3.2} _{-1.9}$ & 0.8 & 5.8 & $^{+3.6} _{-2.3}$ & 0.8 & 4.0 & $^{+3.2} _{-1.9}$ & 0.8 & 14.0 & $^{+5.7} _{-3.6}$ & 1.4 & 15 \\
[\cmsTabSkip]
3600 [3250--5170] & 19.2 & 0.2 & 1.8 & 1.0 & $^{+2.3} _{-0.8}$ & 0.2 & 0.4 & $^{+2.1} _{-0.4}$ & 0.1 & 0.2 & $^{+1.9} _{-0.2}$ & 0.1 & 1.6 & $^{+3.7} _{-0.9}$ & 0.2 & 3.0 \\
[\cmsTabSkip]
\multicolumn{17}{c}{Muon channel}  \\
[\cmsTabSkip]
2200 [1860--2800] & 744.0 & 4.7 & 47.5 & 35.0 & $^{+7.0} _{-5.9}$ & 4.8 & 40.1 & $^{+7.4} _{-6.3}$ & 7.0 & 12.0 & $^{+4.6} _{-3.4}$ & 1.3 & 87.1 & $^{+11.1} _{-9.3}$ & 8.6 & 74 \\
[\cmsTabSkip]
2800 [2430--3930] & 177.0 & 1.1 & 13.1 & 8.4 & $^{+4.0} _{-2.8}$ & 1.3 & 9.9 & $^{+4.3} _{-3.1}$ & 1.8 & 2.7 & $^{+2.8} _{-1.5}$ & 0.3 & 20.9 & $^{+6.5} _{-4.5}$ & 2.2 & 18 \\
[\cmsTabSkip]
3600 [3190--5500] & 29.2 & 0.2 & 2.6 & 1.6 & $^{+2.5} _{-1.1}$ & 0.5 & 0.7 & $^{+2.2} _{-0.7}$ & 0.1 & 0.2 & $^{+1.9} _{-0.2}$ & 0.1 & 2.6 & $^{+3.9} _{-1.3}$ & 0.5 & 4.0 \\
[\cmsTabSkip]
\end{tabular}
}
\end{table}

\begin{figure}[ht]
  \centering
    \includegraphics[width=0.45\textwidth]{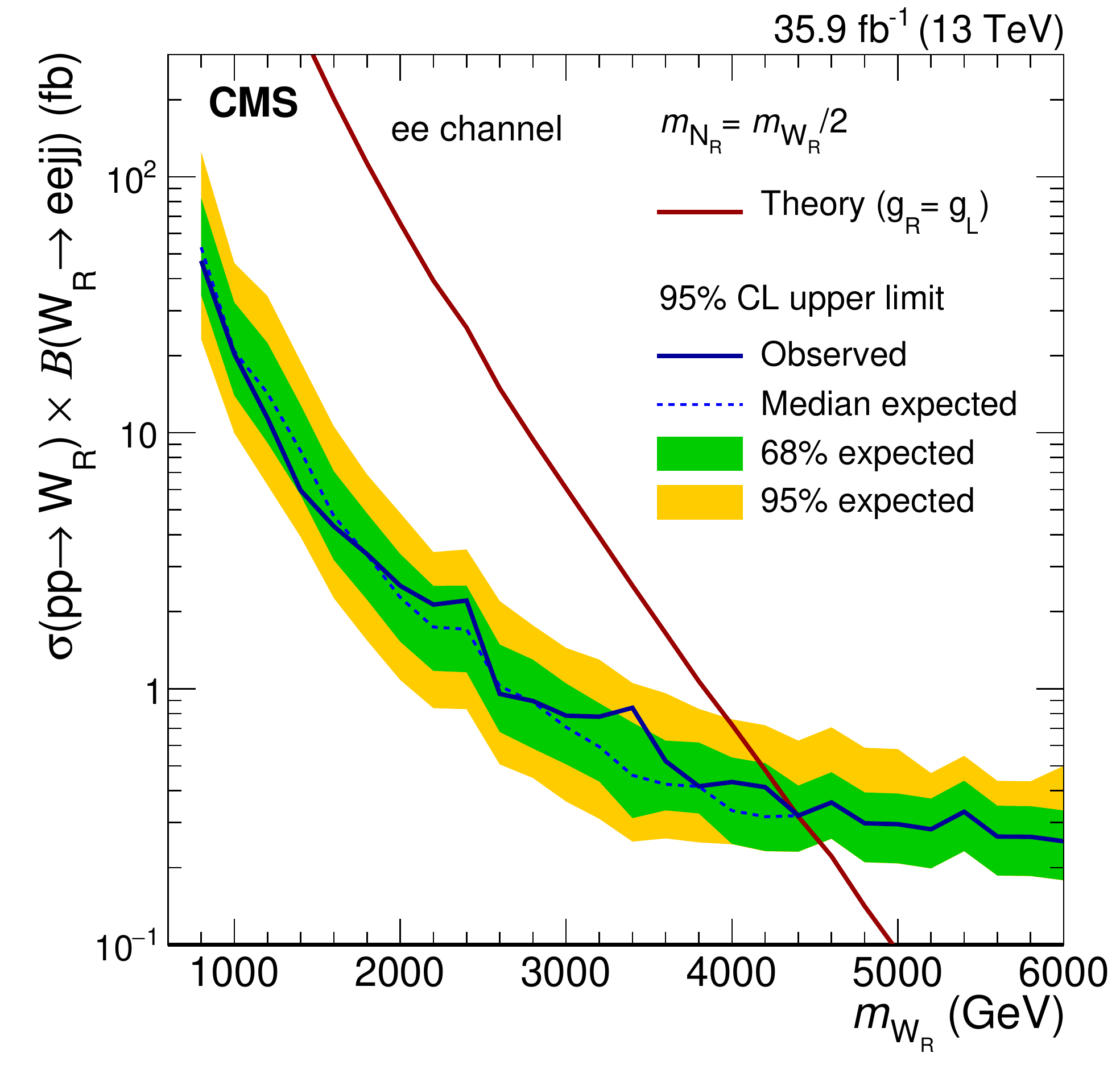}
    \includegraphics[width=0.45\textwidth]{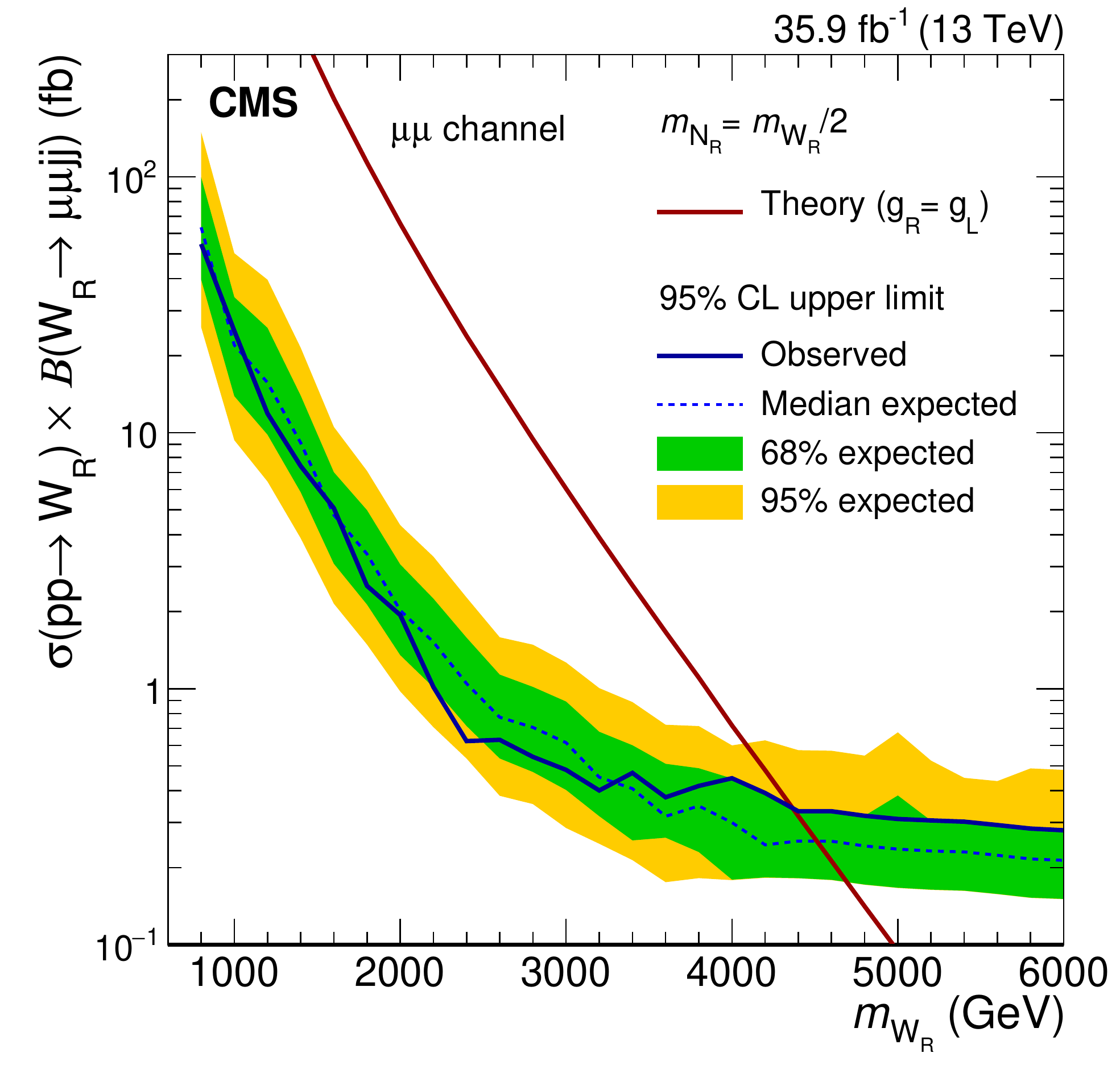}
  \caption{Expected and observed 95\% \CL upper limits on the product of $\sigma(\Pp\Pp \to \WR)$ and branching fraction $B(\WR \to \ell\ell\text{jj})$ for the electron channel on the left and for the muon channel on the right.
  The inner (green) band and the outer (yellow) band indicate the expected 68\% and 95\% \CL exclusion regions.}
  \label{fig:WRxsxnlimwithsystunc}
\end{figure}

Expected and observed exclusion limits on the signal cross section at 95\% \CL are  shown in Fig.~\ref{fig:WRxsxnlimwithsystunc}, taking into account all the systematic and statistical uncertainties described in this section.
For the \WR model, with $\mNR = 1/2 \mWR$, the observed lower limit at 95\% \CL on the mass of the right-handed \PW\ boson is 4.4\TeV for both channels, while the expected exclusion limit is 4.4\TeV for the electron channel and 4.5\TeV for the muon channel, giving an improvement of $\sim$1.4\TeV from the previous analysis at 8\TeV.
The most significant excess, of $\sim$1.5$\sigma$, is observed at $\mlljj\simeq 3.4\TeV$ in the electron channel.
A 2.8$\sigma$ excess seen at $\meejj\approx2.1\TeV$ with the 8\TeV analysis is thus not confirmed by the present data.
The lower edge of the 95\% \CL band disappears at high masses because of the small number of events in that region.
Assuming that only one heavy neutrino flavor \NR contributes significantly to the \WR decay width, the region in the two-dimensional (\mWR, \mNR) mass plane is analyzed, covering a large range of neutrino masses below the \WR boson mass.
The \WR cross section limits obtained for $\mNR=1/2 \mWR$ are scaled to this 2D plane by applying an \mWR- and \mNR-dependent SF to the cross section limit.
This SF is calculated using \WR signal events at the generator level that pass the signal selection, and accounts for the change in the \WR acceptance and efficiency as \mNR\ changes for fixed \mWR.
\begin{figure}[ht]
    \includegraphics[width=0.5\textwidth]{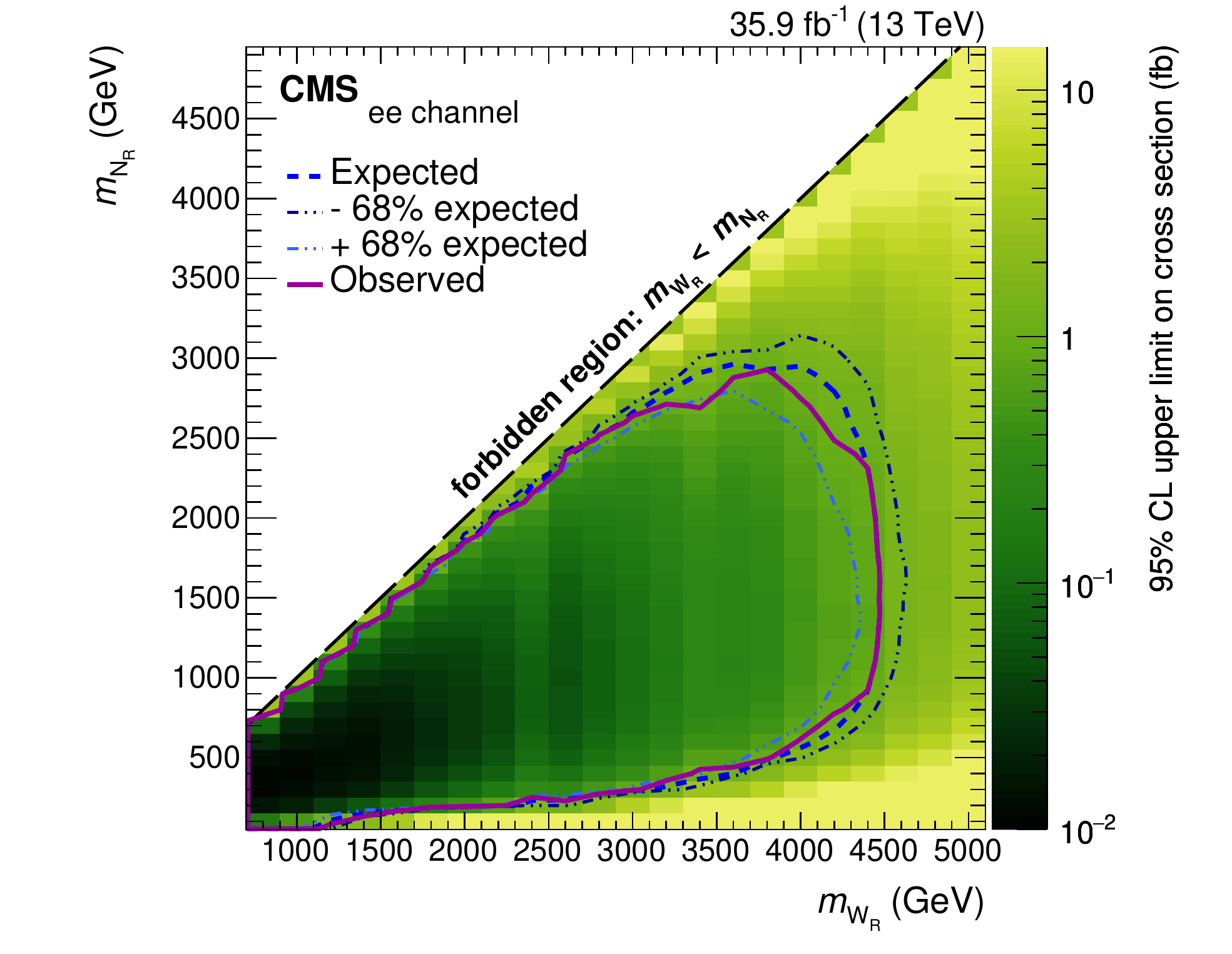}
    \includegraphics[width=0.5\textwidth]{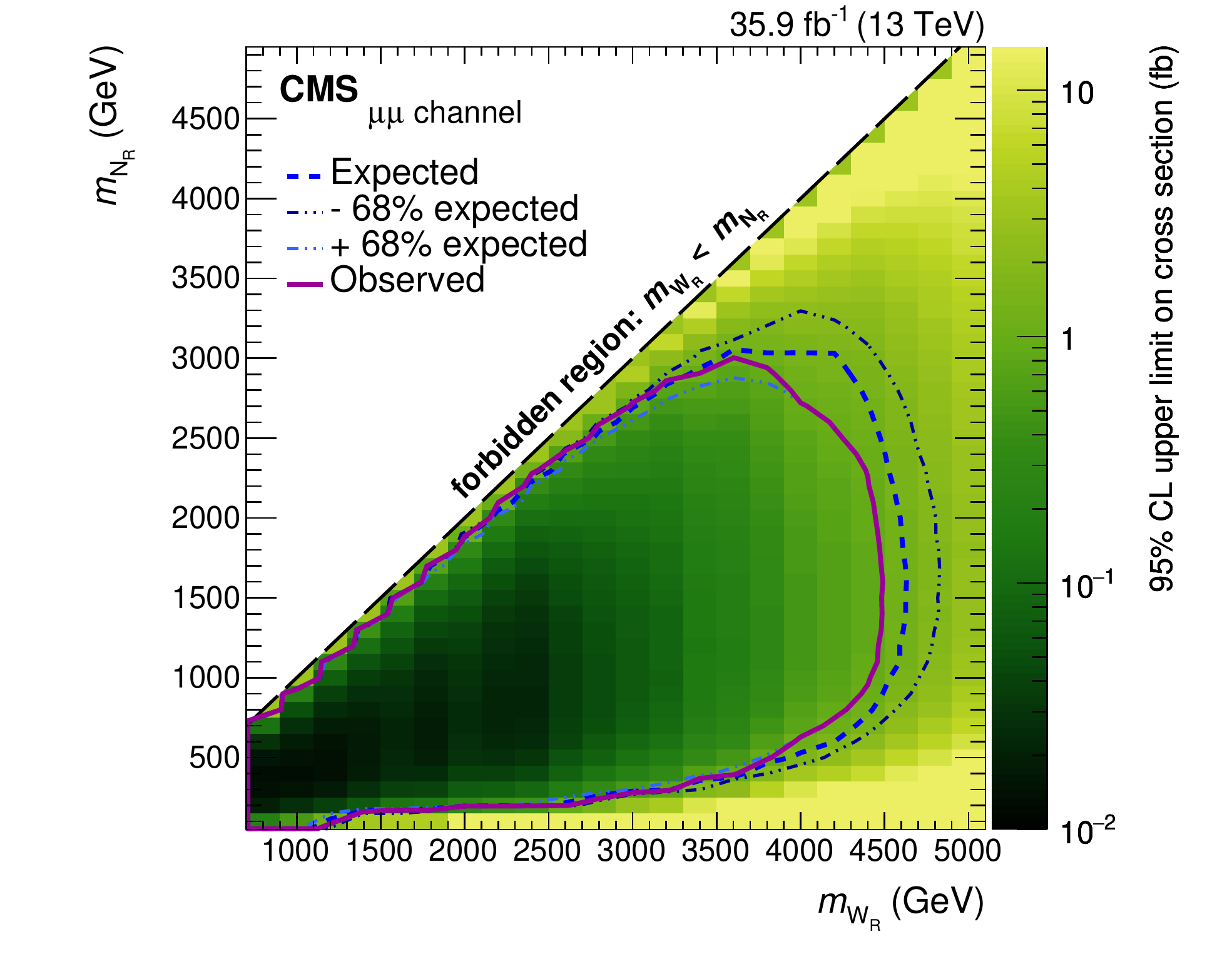}
  \caption{Upper limit on the cross section for different \WR and \NR mass hypotheses, for the electron channel on the left and for the muon channel on the right. The expected and observed exclusions are shown as the dotted (blue) curve and the solid (red) curve, respectively. The thin-dotted (blue) curves indicate the region in (\mWR, \mNR) parameter space that is expected to be excluded at 68\% \CL in the case that no signal is present in the data.
  }
  \label{fig:2DWRxsxnlimwithsystunc}
\end{figure}
The expected and observed upper limits on the cross section for different \WR and \NR mass hypotheses are shown in Fig.~\ref{fig:2DWRxsxnlimwithsystunc}.
The 2D exclusion limits are less stringent in the region $\mNR \lesssim 1/8 \mWR$, where the selection efficiency in generator level events is lower than in fully reconstructed events.

\section{Summary}
\label{sec:conclusion}
A search for a right-handed analogue of the standard model \PW\ boson in the decay channel of two leptons and two jets has been presented.
The analysis is based on proton-proton collision data collected at $\sqrt{s} = 13\TeV$ by the CMS experiment at the LHC in 2016, corresponding to an integrated luminosity of 35.9\fbinv.
No significant excess over the standard model background expectations is observed in the invariant mass distribution of the dilepton plus dijet system.
Thus the 2.8$\sigma$ excess previously observed in data recorded by CMS at 8\TeV is not confirmed.
Assuming that couplings are identical to those of the standard model,
a region in the two-dimensional plane (\mWR, \mNR) covering a large range of right-handed neutrino masses is excluded at 95\% confidence level.
A \WR boson decaying into a right-handed heavy neutrino with a mass $\mNR=1/2 \mWR$ is excluded at 95\% confidence level up to a mass of 4.4\TeV, providing the most stringent limit to date.

\clearpage
\begin{acknowledgments}
We congratulate our colleagues in the CERN accelerator departments for the excellent performance of the LHC and thank the technical and administrative staffs at CERN and at other CMS institutes for their contributions to the success of the CMS effort. In addition, we gratefully acknowledge the computing centers and personnel of the Worldwide LHC Computing Grid for delivering so effectively the computing infrastructure essential to our analyses. Finally, we acknowledge the enduring support for the construction and operation of the LHC and the CMS detector provided by the following funding agencies: BMWFW and FWF (Austria); FNRS and FWO (Belgium); CNPq, CAPES, FAPERJ, and FAPESP (Brazil); MES (Bulgaria); CERN; CAS, MoST, and NSFC (China); COLCIENCIAS (Colombia); MSES and CSF (Croatia); RPF (Cyprus); SENESCYT (Ecuador); MoER, ERC IUT, and ERDF (Estonia); Academy of Finland, MEC, and HIP (Finland); CEA and CNRS/IN2P3 (France); BMBF, DFG, and HGF (Germany); GSRT (Greece); NKFIA (Hungary); DAE and DST (India); IPM (Iran); SFI (Ireland); INFN (Italy); MSIP and NRF (Republic of Korea); LAS (Lithuania); MOE and UM (Malaysia); BUAP, CINVESTAV, CONACYT, LNS, SEP, and UASLP-FAI (Mexico); MBIE (New Zealand); PAEC (Pakistan); MSHE and NSC (Poland); FCT (Portugal); JINR (Dubna); MON, RosAtom, RAS and RFBR (Russia); MESTD (Serbia); SEIDI, CPAN, PCTI and FEDER (Spain); Swiss Funding Agencies (Switzerland); MST (Taipei); ThEPCenter, IPST, STAR, and NSTDA (Thailand); TUBITAK and TAEK (Turkey); NASU and SFFR (Ukraine); STFC (United Kingdom); DOE and NSF (USA).

\hyphenation{Rachada-pisek} Individuals have received support from the Marie-Curie programme and the European Research Council and Horizon 2020 Grant, contract No. 675440 (European Union); the Leventis Foundation; the A. P. Sloan Foundation; the Alexander von Humboldt Foundation; the Belgian Federal Science Policy Office; the Fonds pour la Formation \`a la Recherche dans l'Industrie et dans l'Agriculture (FRIA-Belgium); the Agentschap voor Innovatie door Wetenschap en Technologie (IWT-Belgium); the F.R.S.-FNRS and FWO (Belgium) under the ``Excellence of Science - EOS" - be.h project n. 30820817; the Ministry of Education, Youth and Sports (MEYS) of the Czech Republic; the Lend\"ulet (``Momentum") Programme and the J\'anos Bolyai Research Scholarship of the Hungarian Academy of Sciences, the New National Excellence Program \'UNKP, the NKFIA research grants 123842, 123959, 124845, 124850 and 125105 (Hungary); the Council of Science and Industrial Research, India; the HOMING PLUS programme of the Foundation for Polish Science, cofinanced from European Union, Regional Development Fund, the Mobility Plus programme of the Ministry of Science and Higher Education, the National Science Center (Poland), contracts Harmonia 2014/14/M/ST2/00428, Opus 2014/13/B/ST2/02543, 2014/15/B/ST2/03998, and 2015/19/B/ST2/02861, Sonata-bis 2012/07/E/ST2/01406; the National Priorities Research Program by Qatar National Research Fund; the Programa Estatal de Fomento de la Investigaci{\'o}n Cient{\'i}fica y T{\'e}cnica de Excelencia Mar\'{\i}a de Maeztu, grant MDM-2015-0509 and the Programa Severo Ochoa del Principado de Asturias; the Thalis and Aristeia programmes cofinanced by EU-ESF and the Greek NSRF; the Rachadapisek Sompot Fund for Postdoctoral Fellowship, Chulalongkorn University and the Chulalongkorn Academic into Its 2nd Century Project Advancement Project (Thailand); the Welch Foundation, contract C-1845; and the Weston Havens Foundation (USA).
\end{acknowledgments}

\bibliography{auto_generated}
\cleardoublepage \appendix\section{The CMS Collaboration \label{app:collab}}\begin{sloppypar}\hyphenpenalty=5000\widowpenalty=500\clubpenalty=5000\textbf{Yerevan Physics Institute,  Yerevan,  Armenia}\\*[0pt]
A.M.~Sirunyan, A.~Tumasyan
\vskip\cmsinstskip
\textbf{Institut f\"{u}r Hochenergiephysik,  Wien,  Austria}\\*[0pt]
W.~Adam, F.~Ambrogi, E.~Asilar, T.~Bergauer, J.~Brandstetter, E.~Brondolin, M.~Dragicevic, J.~Er\"{o}, A.~Escalante Del Valle, M.~Flechl, M.~Friedl, R.~Fr\"{u}hwirth\cmsAuthorMark{1}, V.M.~Ghete, J.~Grossmann, J.~Hrubec, M.~Jeitler\cmsAuthorMark{1}, A.~K\"{o}nig, N.~Krammer, I.~Kr\"{a}tschmer, D.~Liko, T.~Madlener, I.~Mikulec, E.~Pree, N.~Rad, H.~Rohringer, J.~Schieck\cmsAuthorMark{1}, R.~Sch\"{o}fbeck, M.~Spanring, D.~Spitzbart, A.~Taurok, W.~Waltenberger, J.~Wittmann, C.-E.~Wulz\cmsAuthorMark{1}, M.~Zarucki
% ...
\vskip\cmsinstskip
% ...
1:~~Also at Vienna University of Technology, Vienna, Austria\\
% ...

\end{sloppypar}
\end{document}